# Extreme nonlinear Raman interaction of an ultrashort nitrogen ion laser with an impulsively excited molecular wavepacket

Zhaoxiang Liu[1,3], Jinping Yao[1,†], Haisu Zhang[2,‡], Bo Xu[1,3], Jinming Chen[1,3,4], Fangbo Zhang[1,3], Zhihao Zhang[1,3,4], Yuexin Wan[1,3], Wei Chu[1], Zhenhua Wang[2], and Ya Cheng[1,2,5,6,*]

[1]*State Key Laboratory of High Field Laser Physics, Shanghai Institute of Optics and Fine Mechanics, Chinese Academy of Sciences, Shanghai 201800, China*

[2]*State Key Laboratory of Precision Spectroscopy, East China Normal University, Shanghai 200062, China*

[3]*University of Chinese Academy of Sciences, Beijing 100049, China*

[4]*School of Physical Science and Technology, ShanghaiTech University, Shanghai 200031, China*

[5]*Collaborative Innovation Center of Extreme Optics, Shanxi University, Taiyuan, Shanxi 030006, China*

[6]*Collaborative Innovation Center of Light Manipulations and Applications, Shandong Normal University, Jinan 250358, China*

[†]*jinpingmrg@163.com*

[‡]*hszhang@phy.ecnu.edu.cn*

[*]*ya.cheng@siom.ac.cn*

**Abstract:**

We report generation of cascaded rotational Raman scattering up to 58$^{th}$ orders in coherently excited $CO_2$ molecules. The high-order Raman scattering, which produces a quasiperiodic frequency comb with more than 600 sidebands, is obtained using an intense femtosecond laser to impulsively excite rotational coherence and the femtosecond-laser-induced $N_2^+$ lasing to generate cascaded Raman signals. The novel configuration allows this experiment to be performed with a single femtosecond laser beam at free-space standoff locations. It is revealed that the efficient spectral extension of Raman signals is attributed to the specific spectra-temporal structures of $N_2^+$ lasing, the ideal spatial overlap of femtosecond laser and $N_2^+$ lasing, and the guiding effect of molecular alignment. The Raman spectrum extending above 2000 cm$^{-1}$ naturally corresponds to a femtosecond pulse train due to the periodic revivals of molecular rotational wavepackets.

Synthesizing femtosecond/subfemtosecond pulses by molecular modulators has been well-established in the recent decades [1]. The highly efficient cascading Raman scattering from molecules brought into concerted vibrations/rotations and hosted in long guiding structures, can easily manipulate the wavelength and broaden the spectrum of incident light while maintains good phase coherence, necessitating a convenient dispersion compensation for ultrashort-pulse generation [4]. Molecular modulators feature a broad operation bandwidth and high energy content at infrared-visible-UV wavelengths, in contrast to the limited bandwidths of usual optoelectronic modulators and the low efficiency of attosecond pulses generated by high-order harmonic generation [7]. Depending on the active Raman frequencies of molecules and the duration of input pulse, both an isolated ultrashort-pulse and a periodic ultrashort-pulse train are generated by adiabatically and/or impulsively prepared molecular modulators [8], holding numerous demanding applications in time-resolved chemistry, optical communications, precision metrology or spectroscopy, to cite a few.

On the other hand, laser-induced molecular alignment has arose increasing interests due the compelling need to manipulate the orientation of molecular axis during diverse physical processes and technological applications. In the nonadiabatic regime, rotationally broad coherent rotational wavepackets are generated in molecules kicked by moderately intense ultrashort laser pulses [12], which manifest themselves as a post-pulse field-free molecular alignment with periodic revivals separated by a full and/or a fraction of the fundamental rotational period. Therefore, the periodic revivals of the rotational wavepackets of constituent molecules can induce an ultrafast modulation of the material refractive index, which has been widely employed to control and localize an incident light pulse both in the spatial and temporal dimensions [[13],[14]].

In this work, we observed multiple high-order rotational Raman sidebands extending above 2000 cm$^{-1}$ in the impulsively excited $CO_2$ molecules, corresponding to a quasiperiodic Raman frequency comb with more than 600 sidebands. A single femtosecond pump laser is utilized to generate the $N_2^+$ lasing as well as the rotational wavepackets in $CO_2$ molecules, which combines the advantages of impulsive preparation of molecular wavepackets with the broadband pump and high spectral resolution with the narrow-bandwidth probe. In particular, the $N_2^+$ lasing served as an ideal light source to generate cascaded Raman scattering from impulsively excited molecules, due to the natural spatial overlap of the lasing radiation with the pump laser and the guided propagation of $N_2^+$ lasing induced by molecular alignment. Moreover, the inherent spectra-temporal structures of the free-space $N_2^+$ lasing favor a high spectral resolution to discriminate individual rotational states and an optimized sampling window of molecular Raman responses [[15][16]]. Specifically, the asymmetric temporal profile of the $N_2^+$ lasing pulse with a fast-rising edge and a slow-trailing edge (as revealed later) is highly desired in fs/ps coherent anti-Stokes Raman scattering (CARS) spectroscopy for gas thermometry and combustion diagnosis [17]. Besides, the few-picoseconds duration of $N_2^+$ lasing greatly mitigates the sensitivity of probe-delays and the group-velocity mismatch induced walk-off between synchronized molecular rotations and the probe pulse. All of the above merits enable, to the best of our knowledge, the first-time observation of cascading rotational Raman scattering as high as the 58th order under free-space conditions. Although one lasing wavelength of $N_2^+$ (i.e. ~428 nm) is selected in the current proof of principle experiment, the versatile wavelength tunability of the $N_2^+$ lasing within the vibrational manifolds of $N_2^+$ electronic transition promises a broad spectral coverage for the attainable Raman frequency combs.

The experiment was performed using a commercial Ti:sapphire laser system (Legend Elite-Duo, Coherent, Inc.), which delivers 800 nm, 40 fs laser pulses with a pulse energy of 6.2 mJ at a repetition rate of 1 kHz. The beam diameter was measured to be about 9 mm. As schematically illustrated in Fig. 1(a), the 800 nm laser beam was focused with an $f$=60 cm lens into a static chamber filled with 15-mbar nitrogen gas. A 30μm-thick beta barium borate (BBO) crystal was used to produce second-harmonic pulses with a central wavelength of 400 nm. In the chamber, the intensity of the 800 nm pulse is measured to $5.9 \times 10^{13}$ W/cm$^2$, whereas the 400 nm pulse is about 2 orders of magnitude weaker. Such a two-color laser field will induce a strong $N_2^+$ lasing radiation at ~428 nm wavelength, as previously reported [[18][19]]. All light signals exiting from the first chamber were collimated and then were focused by an *f*=120 cm lens into the second chamber to generate coherent Raman scattering of $CO_2$ molecules. The Raman scattering signal was collected by a lens into a grating spectrometer with a spectral resolution of 1.6 cm$^{-1}$ (Shamrock 500i, Andor). Additional filters were placed before the spectrometer to improve the signal to noise ratio.

In this experiment, the residual 800 nm laser from the first chamber was used as the pump beam, which impulsively excites rotational coherence of $CO_2$ molecules. Its intensity is estimated to be $1.9 \times 10^{13}$ W/cm$^2$. The $N_2^+$ lasing radiation with the energy of ~0.5 nJ served as the probe beam, which will be inelastically scattered to generate multiple Raman sidebands. As shown in Fig. 1(b), the lasing radiation has a spectral bandwidth of ~2 cm$^{-1}$, enabling the generation of the rotational-state-resolved Raman signal. It also exhibits a good spatial profile and a small divergence angle of 6.2 mrad, allowing a good spatial overlap with the pump beam. The temporal profile of the lasing radiation is further illustrated in Fig. 1(c), featuring a few-picosecond rising edge and a long pulse-wake lasting dozens of picoseconds, as well as a 13 ps delay with respect to the 800 nm pump pulse. All these features inherited from the coherent and superradiant nature of $N_2^+$ lasing as revealed by previous works [[20],[21]].

First, we captured the spectra of coherent Raman scattering at different gas pressures. A series of closely spaced peaks appear on both sides of the $N_2^+$ lasing line, corresponding to the Stokes/anti-Stokes Raman scattering processes. Here, we focus on the Stokes Raman lines lying on the long-wavelength side with corresponding frequency shifts (relative to the $N_2^+$ lasing line) plotted in Fig. 2(a)-(c). At the gas pressure of 20 mbar, as shown in Fig. 2(a), a Gaussian distribution of line-intensities for the rotational Raman sidebands can be clearly observed with frequency shifts determined as $\Delta v = (4J + 6)B$ (green vertical lines), where the rotational constant B=0.3902 cm$^{-1}$ for $CO_2$ molecules in ground-state and *J* is the rotational quantum number [22]. The *J* value corresponding to Raman transition from *J* to *J*+2 state is also indicated. Since only even *J* states are populated for $CO_2$ molecules, the frequency difference between adjacent Raman peaks is a constant value of 8B (i.e., ~3.1 cm$^{-1}$).

With the increase of gas pressures, the second and third bands with larger Raman shifts appear, which correspond to the 2$^{nd}$ and 3$^{rd}$ order rotational Raman scattering, respectively. These high-order Raman scattering signals are generated through cascaded processes, which is schematically illustrated in Fig. 2(d). The 1$^{st}$-order Raman scattering produces a series of Stokes Raman peaks due to the populated multiple rotational states, as observed in Fig. 2(a). Each sideband within the 1$^{st}$ order Raman scattering can act as a new pump source and be further scattered from different initial rotational states to generate multiple 2$^{nd}$ order Stokes components. As a result, the number of Raman sidebands grows with the Raman order. The cascading process is quite efficient due to the pre-excited rotational coherence by the broadband pump laser. Figure 2(e) shows the theoretically calculated spectra of 1$^{st}$, 2$^{nd}$ and 3$^{rd}$ order Raman scattering. The simulation detail is given in the supplemental material [23]. The simulated spectra reproduce main features of the experimental results. Similar to the 1$^{st}$ order Raman scattering, the 2$^{nd}$ and 3$^{rd}$ order Raman scattering spectra are also consisted of equally spaced peaks. For all Raman orders, the strongest Raman peak corresponds to the Raman scattering from $J$=22 to $J$=24. Therefore, two adjacent Raman orders are separated by 94B, as demonstrated experimentally and theoretically. In addition, Fig. 2(e) also clearly shows a broad spectral distribution at a higher Raman order. Spectral overlap among different Raman orders will result in the generation of a flat structure in the total Raman spectrum, as observed in Fig. 2(c). Based on these spectral analyses, we confirmed the presence of high-order cascaded Raman scattering from $CO_2$ molecules.

When the gas pressure is further increased, higher order Raman signals are observed. As illustrated in Fig. 3, Raman scattering can be extended to 17$^{th}$, 32$^{th}$, 46$^{th}$ and 58$^{th}$ orders at the gas pressure of 150 mbar, 300 mbar, 500 mbar and 1 atm, respectively. Such a high-order cascaded Raman scattering process induces a frequency shift >2000 cm$^{-1}$ in the 1-atm $CO_2$ gas, where the conversion efficiency from the $N_2^+$ lasing to all Raman sidebands is estimated to be about 9%. The Raman signal propagates along the direction of the pump laser with a divergence angle of 8.5 mrad. Its spatial profile in the far field is shown in inset of Fig. 3. These characteristics further demonstrate its nature of coherent Raman scattering. More interestingly, unlike most cascaded nonlinear processes, Raman signals do not show a sharp decay with the increase of Raman order. Instead, it exhibits a plateau in the high-order Raman spectra, which is particularly obvious at 1atm. In addition, two strong spectral lines with the frequency shift of 1285 cm$^{-1}$ and 1388 cm$^{-1}$ are overlaid on the rotational Raman spectra, which correspond vibrational Raman scatterings of $02^00(\Sigma_g^+)$ and $10^00(\Sigma_g^+)$ states of $CO_2$ molecules, respectively [22].

Furthermore, we performed a cross-correlation measurement to obtain the temporal information of these high-order Raman scattering signals. To this end, a weak 800 nm laser beam together with the Raman signal generated at 1 atm are launched into a BBO crystal to generate the time- and frequency-resolved sum frequency signal (SFS). After integrating spectrally, we can obtain the SFS as a function of the time delay of two beams, which approximately reflects the temporal structure of Raman scattering. As indicated by the red dot-dash line in Fig. 4, Raman signal is composed of a pulse train with a constant interval of ~10.7 ps, which approximately equals one quarter of the full rotational period of $CO_2$ molecules $T_{rev}$. The measured result clearly shows that strong Raman scattering only occurs at the revival moments of rotational wavepackets. The relative intensity of these sub-pulses basically follows the temporal envelop of the $N_2^+$ lasing. In this perspective, the $N_2^+$ lasing signal with picoseconds duration is transformed into a femtosecond pulse-train through coherent cascaded rotational Raman scattering. Previous studies show that both the pulse duration of $N_2^+$ lasing and its delay with respect to the 400 nm seed pulse strongly depend on gas pressure [[20],[21]]. Therefore, by simply adjusting the gas pressure in the lasing generation chamber, active control of the spectra-temporal profiles of $N_2^+$ lasing can be realized and further used to engineer the high-order cascaded Raman spectrum as well as the temporal profile of the synthesized pulse.

To gain a deep understanding on the temporal information, we calculated the coherent Raman scattering by solving the simplified nonlinear wave equation [[24]]

$$\left(\frac{\partial^2}{\partial z^2} - \frac{n_0^2}{c^2}\frac{\partial^2}{\partial t^2}\right)E_p = \frac{4\pi}{c^2}\frac{\partial^2}{\partial t^2}P^{(3)}. \tag{1}$$

Here, $P^{(3)} = \chi^{(3)}|E_i|^2 E_p$ is the nonlinear polarization to describe the coherent Raman scattering with the third-order susceptibility $\chi^{(3)}$ and the pump (probe) laser field $E_i$ ($E_p$). By solving Eq. (1) with the slowly varying envelope approximation, we obtain

$$E_p(z,t) = |A_0(t-\tau)|\cos[\omega_0(t-\tau) - kz], \tag{2}$$

where $k = \frac{\omega_0}{c}n(t)$, $\tau$ is the time delay between the pump and probe pulses, $A_0$ and $\omega_0$ are the temporal envelope and the carrier frequency of the probe field, respectively. Since polarizations of the pump and probe laser fields are nearly parallel to each other in this experiment, the temporally modulated refractive index $n(t)$ is given by [[10],[13]]

$$n(t) \approx 1 + \frac{N}{2\varepsilon_0}[\alpha_\perp + (\alpha_\parallel - \alpha_\perp)\langle\cos^2(\theta)\rangle(t)]. \tag{3}$$

Here, $N$ is the number density of $CO_2$ molecules, $\alpha_\parallel$ ($\alpha_\perp$) is the parallel (perpendicular) component of the anisotropic polarizability at the probe wavelength, $\varepsilon_0$ is the electric constant, and $\langle\cos^2(\theta)\rangle(t)$ is the alignment factor determined by the pump field with $\theta$ being the angle of molecular axis with respect to the pump laser field. The alignment factor is calculated by solving the time-dependent Schrödinger equation based on the rigid-rotor model [[25],[26]]. Furthermore, we perform the time-frequency analyses on the probe laser field in Eq. (2) using Wavelet transformation, and the temporal profile of the Raman scattering signal is obtained by spectral integration and illustrated as green solid line in Fig. 4 (corresponding detail is given in the supplemental material [23]). Similar to the experimental observation, Raman scattering efficiently occurs near the rotational revivals where the maximum gradient of refractive index emerges (also see Fig. R2 of supplemental material).

It is noteworthy that the observed high-order Raman scattering is achieved in free-space conditions, in contrary to usually utilized meter-scale gas-filled hollow-core waveguides/fibers for efficient cascading Raman scattering generation in the nanosecond pumping scheme [[5],[6]]. This is because the high intensity of femtosecond laser pulses enables us to generate a channel of rotational quantum wavepackets with a length much longer than Rayleigh range, as discussed in the supplemental material [23]. Furthermore, the guiding effect of the femtosecond-laser-induced rotational wavepackets further promotes the interaction of $N_2^+$ lasing with the Raman active medium. Briefly speaking, at revival moments of the rotational wavepackets, an increase of refractive index in the center compared to the surrounding region behaves like a positive lens to guide the diffracted and scattered $N_2^+$ lasing radiation propagating through the channel of rotational wavepackets. In addition, experimental results show that Raman scattering can be further extended to higher order by employing the stronger pump laser, the looser focusing geometry and the higher gas pressure [23].

To conclude, we have demonstrated the generation of multiple cascaded rotational Raman scattering as high as the 58th order, by the interaction of $N_2^+$ lasing radiation with impulsively excited $CO_2$ molecules. The Raman scattering spectrum is composed of a broad quasiperiodic frequency comb with more than 600 sidebands, which naturally corresponds to a femtosecond pulse train in the temporal domain. The experimental and theoretical analyses reveal that such a high-order cascaded Raman scattering is attributed to the specific time-frequency and spatial characteristics of $N_2^+$ lasing as well as the guiding effect of molecular alignment. The former allows for a good spatio-temporal coupling between the pre-excited molecular wavepackets and probe pulses, while the latter ensures a long Raman active medium. This work opens promising applications of air lasing on ultrafast nonlinear spectroscopy and optical frequency comb generation.

This work is supported by the National Key Research and Development Program of

**References:**

[1] S. Baker, I. A. Walmsley, J. W. G. Tisch, and J. P. Marangos, Nat. Photon. **5**, 664 (2011).

[2] H. S. Chan, Z. M. Hsieh, W. H. Liang, A. H. Kung, C. K. Lee, C. J. Lai, R. P. Pan, and L. H. Peng, Science **331**, 1165 (2011).

[3] S. E. Harris and A. V. Sokolov, Phys. Rev. Lett. **81**, 2894 (1998).

[4] A. V. Sokolov, D. R. Walker, D. D. Yavuz, G. Y. Yin, and S. E. Harris, Phys. Rev. Lett. **87**, 033402 (2001).

[5] F. Couny, F. Benabid, P. J. Roberts, P. S. Light, and M. G. Raymer, Science **318**, 1118 (2007).

[6] P. Hosseini, A. Abdolvand, and P. St. J. Russell, Opt. Lett. **41**, 5543 (2016).

[7] F. Krausz and M. Ivanov, Rev. Mod. Phys. **81**, 163 (2009).

[8] N. Zhavoronkov and G. Korn, Phys. Rev. Lett. **88**, 203901 (2002).

[9] D. D. Yavuz, D. R. Walker, M. Y. Shverdin, G. Y. Yin, and S. E. Harris, Phys. Rev. Lett. **91**, 233602 (2003).

[10] R. A. Bartels, T. C. Weinacht, N. Wagner, M. Baertschy, C. H. Greene, M. M. Murnane, and H. C. Kapteyn, Phys. Rev. Lett. **88**, 013903 (2002).

[11] M. Wittmann, A. Nazarkin, and G. Korn, Opt. Lett. **26**, 298 (2001).

[12] H. Stapelfeldt, and T. Seideman, Rev. Mod. Phys. **75**, 543 (2003).

[13] F. Calegari, C. Vozzi, S. Gasilov, E. Benedetti, G. Sansone, M. Nisoli, S. De Silvestri, and S. Stagira, Phys. Rev. Lett. **100**, 123006 (2008).

[14] S. Varma, Y.-H. Chen, and H. M. Milchberg, Phys. Rev. Lett. **101**, 205001 (2008).

[15] D. Pestov, R. K. Murawski, G. O. Ariunbold, X. Wang, M. Zhi, A. V. Sokolov, V. A. Sautenkov, Y. V. Rostovtsev, A. Dogariu, Y. Huang, and M. O. Scully,

Science **316**, 265 (2007).

[16] S. P. Kearney, Combust. Flame **162**, 1748 (2015).

[17] H. U. Stauffer, J. D. Miller, S. Roy, J. R. Gord, and T. R. Meyer, J. Chem. Phys. **136**, 111101 (2012).

[18] J. Yao, G. Li, C. Jing, B. Zeng, W. Chu, J. Ni, H. Zhang, H. Xie, C. Zhang, H. Li, H. Xu, S. L. Chin, Y. Cheng, and Z. Xu, New J. Phys. **15**, 023046 (2013).

[19] B. Xu, J. Yao, Y. Wan, J. Chen, Z. Liu, F. Zhang, W. Chu, and Y. Cheng, Opt. Express, **27**, 18262 (2019).

[20] G. Li, C. Jing, B. Zeng, H. Xie, J. Yao, W. Chu, J. Ni, H. Zhang, H. Xu, Y. Cheng, and Z. Xu, Phys. Rev. A **89**, 033833 (2014).

[21] Y. Liu, P. Ding, G. Lambert, A. Houard, V. Tikhonchuk, and A. Mysyrowicz, Phys. Rev. Lett. **115**, 133203 (2015).

[22] G. Herzberg, and L. Herzberg, J. Opt. Soc. Am. **43**, 1037 (1953).

[23] See Supplemental Material for the details on the simulated Raman scattering spectra, time-frequency analyses on cascaded Raman scattering, and Raman shift as functions of the interaction length and the pump energy.

[24] R. W. Boyd, *Nonlinear Optics*, 3rd ed. (Elsevier, Singapore, 2008).

[25] T. Seideman, J. Chem. Phys. **103**, 7887 (1995).

[26] J. Ortigoso, M. Rodríguez, M. Gupta, and B. Friedrich, J. Chem. Phys. **110**, 3870 (1999).

**Captions of figures:**

Fig. 1 (Color online) (a) Schematic of the experimental setup for $N_2^+$ lasing generation and coherent Raman scattering. (b) The measured spectrum of $N_2^+$ lasing and its spatial profile (inset). (c) The temporal information of $N_2^+$ lasing (blue solid line) and the 400 nm laser pulses (black dot line) obtained by the cross-correlation technique. In this measurement, the sum frequency of the $N_2^+$ lasing signal or the 400 nm pulses and a weak 800 nm pulse in a BBO crystal is recorded as a function of the delay. The zero delay is defined with the peak of the 800 nm laser pulse, as indicated with a red arrow.

Fig. 2 (Color online) Measured rotational Raman spectra (blue solid curves) at the gas pressure of (a) 20 mbar, (b) 60 mbar, and (c) 120 mbar, respectively. The theoretically calculated Raman shifts are indicated by vertical lines. (d) Schematic diagram of the first, second and third order rotational Raman scattering. (e) Simulated Raman spectra for the first three order Raman scattering. Each order Raman scattering is normalized by itself.

Fig. 3 (Color online) The measured Raman spectra at the gas pressure of 150 mbar (green line), 300 mbar (purple line), 500 mbar (red line) and 1 atm (blue line), respectively. The inset shows the spatial profile of Raman signals at 1 atm.

Fig. 4 (Color online) Temporal structures of Raman scattering at 1 atm (red dot-dash line) and $N_2^+$ lasing (blue solid line) measured with the cross-correlation technique. The simulated temporal profile of Raman scattering is indicated with the green solid line.

Fig. 1

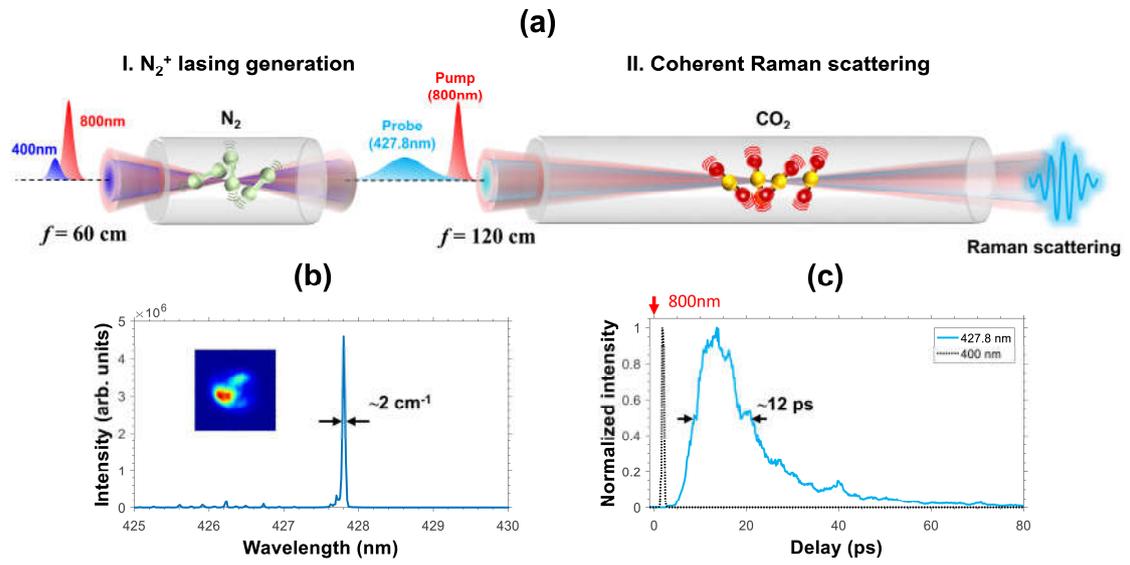

Fig. 2

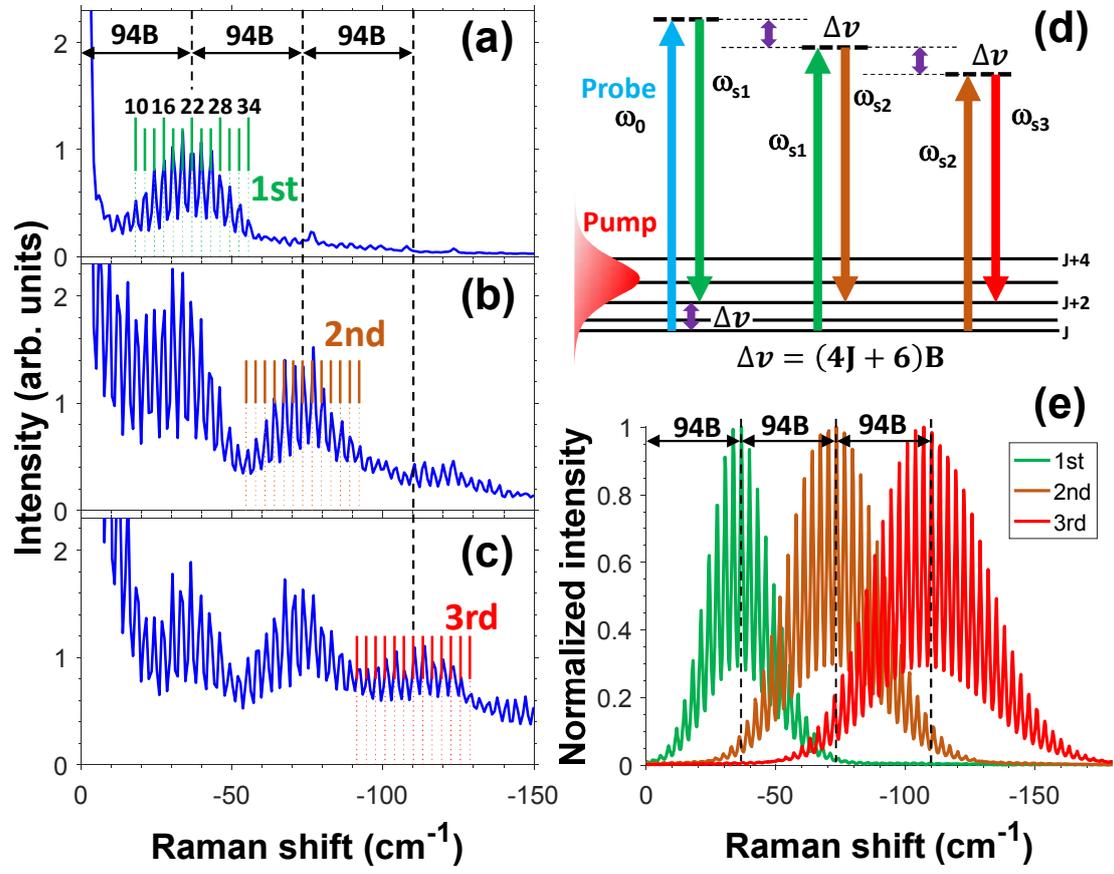

Fig. 3

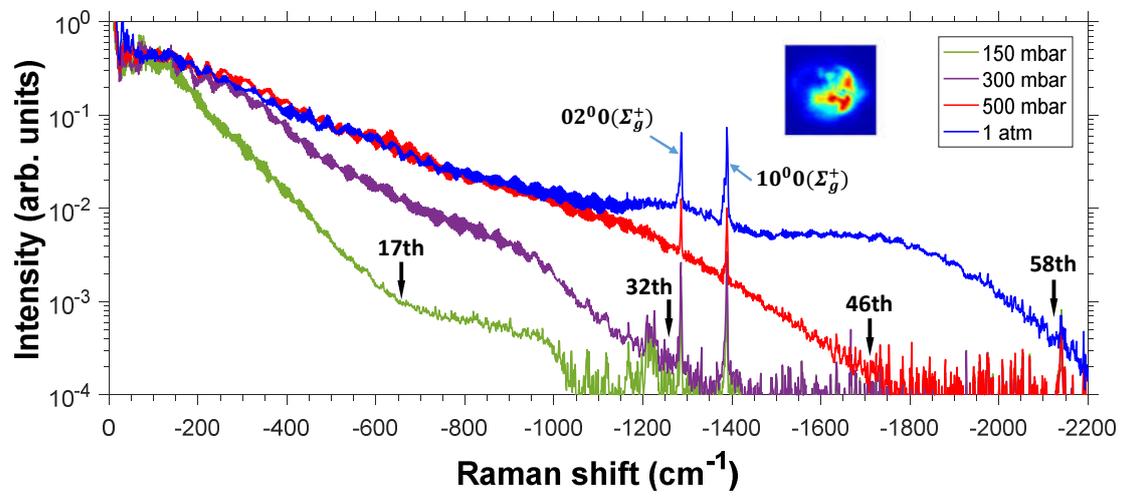

Fig. 4

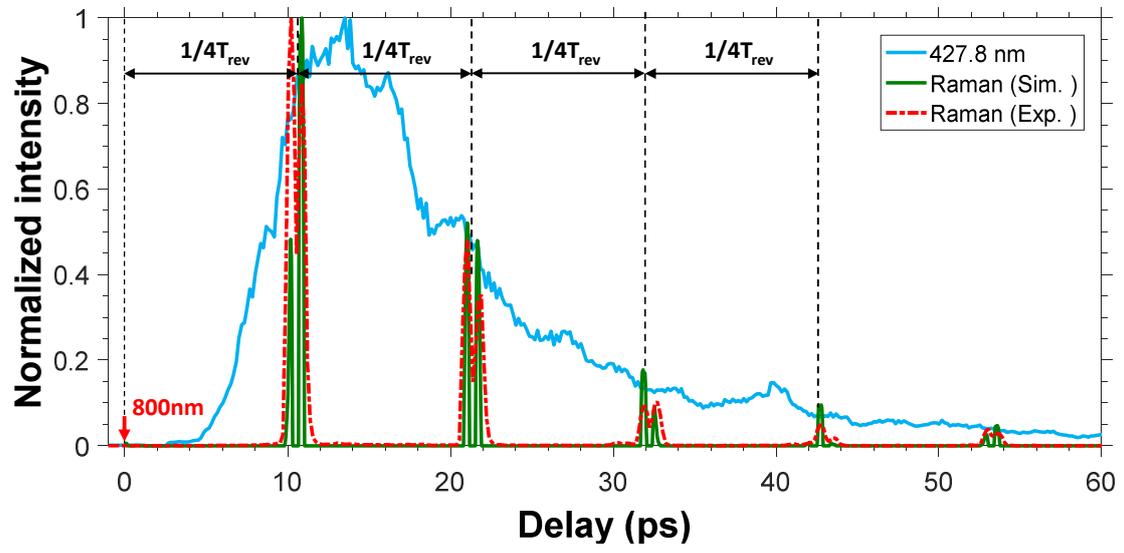